\documentclass{article}

\usepackage{graphicx}

\usepackage{subcaption}

\begin{document}
\def\boxit#1{\vcenter{\hrule\hbox{\vrule\kern8pt
      \vbox{\kern8pt#1\kern8pt}\kern8pt\vrule}\hrule}}
\def\Boxed#1{\boxit{\hbox{$\displaystyle{#1}$}}} 
\def\sqr#1#2{{\vcenter{\vbox{\hrule height.#2pt
        \hbox{\vrule width.#2pt height#1pt \kern#1pt
          \vrule width.#2pt}
        \hrule height.#2pt}}}}
\def\square{\mathchoice\sqr34\sqr34\sqr{2.1}3\sqr{1.5}3}
\def\Square{\mathchoice\sqr67\sqr67\sqr{5.1}3\sqr{1.5}3}
\def\AJP{{\it Am. J. Phys.}}
\def\AM{{\it Ann. Math.}}
\def\AP{{\it Ann. Phys.}}
\def\CQG{{\it Class. Quantum Grav.}}
\def\GRG{{\it Gen. Rel. Grav.}}
\def\JMP{{\it J. Math. Phys.}}
\def\JP{{\it J. Phys.}}
\def\JSIRAN{{\it J. Sci. I. R. Iran}}
\def\NC{{\it Nuovo Cim.}}
\def\NP{{\it Nucl. Phys.}}
\def\PL{{\it Phys. Lett.}}
\def\PR{{\it Phys. Rev.}}
\def\PRL{{\it Phys. Rev. Lett.}}
\def\PRp{{\it Phys. Rep.}}
\def\RMP{{\it Rev. Mod. Phys.}}
\title{\bf Magnetic Dipole and Noncommutativity}
\author{{\small Behrooz Malekolkalami}\footnote{b.malakolkalami@uok.ac.ir}, {\small Taimur Mohammadi}\footnote{t.mohammadi@uast.ac.ir}\\
        {\small Department of Physics,  University of Kurdistan, Pasdaran St., }\\
        {\small Sanandaj, Iran}}
\date{\small \today}
\maketitle
\indent
\begin{abstract}
The noncommutativity concept has wide range of applications in physical and mathematical theories.
Noncommutativity in the position--time coordinates  concerns the microscale  structure of space--time. The noncommutativity is  an intrinsic propery  of the space--time and  it could be different from usual properties when one encounters  the high energy  phenomena.  On the other hand,  the space--time is assumed to be  as a background for the occurrence of physical events.  Therefore, it is not far--fetched to expect the emergence of new   physics or dynamics  when the fine geometric structure of  space--time is deformed. In this work, we consider a common form of this deformation and try to  answer the question as: a physical (or dynamical) model can be described by the noncommutative effects?. This  can also be asked this way:  does the noncommutativity could have a physical manifestations in the nature? Our model here is a magnetic dipole.
\end{abstract}
\medskip
{\small \noindent
\bigskip
\section{Introduction}
\indent
The Noncommutative (\textbf{NC}) geometry has a relatively long history in mathematics
and physics  and  has had an increasingly important role in the attempts to understand the space--time structure at the microscale  scales. The NC  idea at the  micro distances, first introduced by Snyder \cite{Snyder}. He applied the concept of NC structure to discrete space--time coordinates instead of continuous ones. The NC geometry provides valuable tools for the study of physical theories in classical and quantum level. In recently decades, the NC version of physical theories  has been widely employed in many of the research works. The main motivations for interest in NC field theories come from string theory and quantum gravity \cite{Seiberg}.\\
The noncommutativity\footnote{In short, it is sometimes referred to as deformation.} in  space--time coordinates has   kinematic effects, because  kinematics deals with the all possible motion of material systems  and their trajectories based on geometric description which concerns the  structure of the configuration space. In other words, the noncommutativity can be regarded as a type of   deformation of space structure. However in order to include  the effects of noncommutativity in classical and quantum dynamics, the deformation should also contain the phase space as a whole (geometric) structure. Routine to describe the deformation is  change in the symplectic structure\rlap.\footnote{It is easily to show that, An inherent definition of  \textit{Poisson Brackets} can be  given within that framework where  the symplectic structure is outlined, see  e. g. \cite{Gosson} (page 11).} For example, when the noncommutativity is  described through the  NC parameters ($\alpha_{ij}$) and ($\beta_{ij}$), corresponding to the space and momentum sectors of the phase space, respectively, one realizes that the definitions of parameters are given by the \emph{Poisson Brackets} which in turn are derived from the corresponding symplectic structure. A brief description of this subject may be useful:\\
The \emph{Generalized Poisson Brackets} of two (dynamical) variables ($A, B$) is defined as
\begin {center}$\{A, B\}_{\omega}=\omega(A, B)=\omega^{ij}\frac{\partial A}{\partial x^i}\frac{\partial B}{\partial x^j},$\end{center}
which  $\omega$ is the given symplectic structure  and $\omega_{ij}$ is the entries  of (corresponding) symplectic matrix.
The time evolution of any dynamical variables,  say $f$  is also given by
\begin {center}$\dot{f}=\{H, f\}_{\omega}=\omega(H, f)$,\end{center}
where $H$ is the Hamiltonian. For the  phase space variables, say $x=(q, p)$, the  Hamilton equations are
\begin{equation}
 \dot{x}_i=\{H, x_i\}_{\omega}=\omega(H, x_i)=\omega^{ij}\frac{\partial H}{\partial x^j}.
\end{equation}
In the usual phase space (commutative case), the values of  the symplectic matrix elements are merely the numbers $\omega_{ij}= -1, 0, 1$, however in  NC phase space, $\omega_{ij}$ is in terms of the NC parameters and hence the NC Hamilton equations contain  additional terms  with respect to the usual (commutative) case.  These additional  terms  are (new) forces arisen from the noncommutativity and the interpretation of them may lead to familiar or new physics. We refer to this as equivalency between  noncommutative effects  and  physical  phenomena.

The examples of such  equivalencies have been demonstrated in the classical and the quantum systems. As first example, it is possible to consider a new form of NC phase space with an operatorial form of noncommutativity and  showing that  a moving particle  feels the effect of an interaction with an effective magnetic field \cite{Ghosh}. Another example is that a  term appearing in the  NC equations of motion (including NC parameter   $\beta_{ij}$)  behaves as a constant magnetic force \cite{Djamai}. More precisely, a moving free particle in NC space is acted upon by a force similar to that one acting upon  a moving charged particle in a (constant) magnetic field in the usual space. Or in other words, the effect of the NC parameter (of the momentum sector  $\beta_{ij}$) in the NC motion equations is analogous to the presence of a constant magnetic force.\\
Here it is necessary to remind that the constant magnetic field in \cite{Djamai}  can  be canceled out by the Lorentz transformations. It depends on the \emph{Inertial Frame of Observer}, what  is sometimes called the \emph{Extrinsic Magnetic Field}. Such an magentic field could be generated, for example,  by an (infinity extension) surface current of constant density and hence in the reference frame of current velocity, it disappears.

In connection with the present work, we are dealing with the \emph{Intrinsic Magnetic Field}, the field that can not be removed by the Lorentz   transformations. It could be generated  by any quasi circuit current, for example, a charged rigid body which rotates around  an axis  passing through the body's center (e. g.  a  solid charged spinning sphere). In electrodynamics, the magnetic field of such body is  approximated (outside the body) by the  magnetic dipole field as the distance from the body increases (or its dimensions are reduced to zero).

Here,  we are going to study and investigate the  motion equations of a test particle moving  in  surrounding space of the  rotating charged sphere in the usual space and compare them  with ones come from the case of a static charged sphere in the NC space. The latter case is a Keplerian type potential which  has been studied in literature, e. g. \cite{Romero, Romero1, Mirza}. One of the main objectives in the present  work  is the approach  to show that the noncommutativity (effects)  can be equivalent to presence of intrinsic magnetic field (or vise versa). A familiar (or  perhaps the most important) example is the magnetic field of a magnetic dipole . The approach can be summarized  in the following paragraph:

Consider a test particle moving in surrounding space of a charged static sphere (as a Coulomb  potential source) in the NC space and a  test particle moving in (far away) outside of a spinning charged sphere in the usual space. By comparing the motion equations in the both spaces, one   realize that, the force term  arisen from noncommutativity  (in the NC space) is analogous to  the magnetic force arisen from the magnetic dipole  in the usual space. In other words, in the NC space, the noncommutativity plays the role of the  rotation.\\
The importance of this case (spinning charged sphere) may be due to the fact that  the magnetic field around any magnetic source looks increasingly like the field of a magnetic dipole as the distance from the source increases (or the  dimensions of the source are reduced to zero what it referred to as \emph{dipole approximation}). On the other hand, as we know, the basic unit of  electrostatic is the electric  monopole (elementary charge), however in  the magnetism, there is not any monopole source for the magnetic field. So far the existence of magnetic monopoles as isolated magnetic (charge) has not been established, but the magnetic dipole is a good alternative to the monopole  in magnetism. Since  the magnetic dipoles experience torque in the presence of magnetic fields, the measurements of  magnetic field can  be done through  the torque which is exerted on a given magnetic source by a certain known dipole magnetic moment.  Hence,  this  basic magnetic unite  has a central and fundamental role in the magnetic interactions with a considerable practical relevance. This is why the magnetic dipoles has numerous  applications in many branches of physics and other natural sciences. For example, magnetic moment of the nucleus in nuclear interactions, or applications of magnetic moment  in molecular magnetism and applied engineering sciences. So, the study of this basic magnetic unit and its sources  can be of particular importance. Our attempt here is that to attribute such  magnetic sources to the geometric properties of   the space,  equivalently, exploring a geometric source for  the  magnetic dipole as  a purely physical identity. Here,  it is appropriate  to note that our approach in this work is by considering the noncommutativity in the  classical phase space, however,   the noncommutativity can be  generalized to the  phase space of  quantum  variables \cite{Nikita}. Furthermore, in mathematical physics,  the NC field theory is an application of noncommutativity to the phase space of  field theories. Accordingly, by constructing  the NC field theory\footnote{In which the phase space variables, say the two operator $X_{\mu}, X_{\nu}$ obey the algebraic relation $[X_{\mu}, X_{\nu}]=i \theta_{\mu \nu}$. }   in the gauge covariantly extending field equations, it can be shown the electric and magnetic (dipole) fields can be produced by an extended static charge \cite{Adorno}.\\
We close this section with a few remarks on  gravitomagnetism. The similarity between electromagnetism and gravitomagnetism brings to mind that the similar equivalencies may exist for gravitomagnetic effects.  These equivalencies have been shown in two previous works, namely equivalence between extrinsic gravitomagnetic field and  noncommmutativity (corresponding to momentum sector)\cite{extrinsic} and equivalence between intrinsic gravitomagnetic field and  noncommmutativity (corresponding to space sector)\cite{intrinsic}. Remarkably,  despite of similarities between electromagnetism and gravitomagnetism  which lead to the some calculation analogies between the present work and \cite{intrinsic}, the discussions here contain new points and have its own specific results which refers to the inherent differences between electromagnetism and gravitomagnetism. The differences include both practical and theoretical which we have listed  some of the most important ones below:\\
1- The gravitational interactions are governed by the \emph{Equivalence Principle} which states that, these interactions   can always be locally eliminated, or are locally indistinguishable from the effects of an accelerated frame. However, in electromagnetic, there is no such principle.  This means that, for example, the two test charges with different electric charges are differently affected by the same external field and  introducing an accelerated
frame  can  locally eliminate  the electromagnetic force acting on  one of the charges only, but not both of them. So the electromagnetic interactions can not be eliminated locally.\\
2- The gravity is a (weak) force with only one sign of mass charge, while opposite poles of two small bar magnets attract each other with a much stronger force than their mutual gravitational attraction. This shows the   force of gravity is negligible  compared to the magnetic force.\\
3- The mass charge  is equivalent to inertia, against the electric charge which is unrelated to inertia.\\
4- The gravitation is a one spin  theory and electromagnetism is a spin half theory.\\
As mentioned, here is what distinguishes our discussion  is due to the substantive  differences  between electromagnetism and gravitomagnetism. For example, the second difference makes the some basic dynamical equations (in this work) include the mass of test particle while their gravitational counterparts in  \cite{intrinsic}  are independent from the mass. Also, remember that the gravitational fields (interactions) are  much weaker than the electromagnetic ones and why is it very difficult to measure and detect. As far as we are know, the  experimental tests of these interaction has been almost impossible in terrestrial laboratories  against the electromagnetic effects which as far as current experimental results show they are easy to measure and detect. So, from the practical point of view,  there is a \emph{particular preference} of the electromagnetic interactions  over the gravitomagnetic ones for the \emph{experimental verification of noncommutativity.} \vspace{3mm}\\
The work is organized as follows:\\
Section 2 gives a brief overview  of NC classical mechanics essential for the work. In section 3,  the equations of motion for a test particle moving in the NC and usual spaces are introduced. Section 4 demonstrates the  equivalence relation between  NC effects and intrinsic magnetic field.  A general discussion  and results is presented in Section 5. Conclusions  are given in the last section.\vspace{1 cm}
\section{Noncommutative Classical Mechanics}
The noncommutativity  concept has its  roots in quantum theory  due to the algebraic structure of quantum  mechanics. The non--commuting operators appear naturally in quantum mechanics. Usual quantum mechanics formulated on commutative space satisfying the following commutation relations:
\begin{equation}
[\hat{x}_{i}, \hat{x}_{j}]=0, \quad[\hat{x}_{i}, \hat{p}_{j}]=i\hbar\delta_{ij}, \quad [\hat{p}_{i}, \hat{p}_{j}]=0,
\end{equation}
where $\hat{x}_{i}$ and $\hat{p}_{j}$ denote coordinates and momentum operators, respectively.
From the mathematical  point of view, the coordinates operators of  NC space--time do not satisfy the usual commutation
relations, and instead obey non--trivial ones, e. g.  $[\hat{x}_{i}, \hat{x}_{j}]\neq 0$. The commutation
relations are defined with respect to the notion of $\ast$ product (read  NC product) such that, generally,  for  two
different operators, say $\hat{A}$ and $\hat{B}$, the inequality   $\hat{A}\ast\hat{B}\neq \hat{B}\ast\hat{A}$ holds. The NC product determines the symplectic structure of the phase space. Depending on symplectic structure,  a variety of NC phase spaces can emerge. Let us consider the simplest  NC quantum mechanics satisfying the  relations:
\begin{equation}
[\hat{x}_{i}, \hat{x}_{j}]=i\hbar\alpha_{ij}\quad [\hat{x}_{i}, \hat{p}_{j}]=i\hbar\delta_{ij}, \quad [\hat{p}_{i}, \hat{p}_{j}]=i\hbar\beta_{ij},
\end{equation}
where $\alpha_{ij}$ and $\beta_{ij}$ are the NC parameters corresponding to the space and momentum sector, respectively. They are assumed to be  real constants and anti--symmetric c--numberc\rlap.\footnote{The anti--symmetric property is usually defined in terms of Levi-Civita symbol $\epsilon_{ijk}$, that is $\alpha_{ij}=\epsilon_{ijk}\alpha_{k}$ and $\beta_{ij}=\epsilon_{ijk}\beta_{k}$.}
The classical mechanics is thought of as a suitable limit of quantum mechanics and transition from quantum mechanics to classical mechanics is facilitated by  the \emph{Dirac Quantization Condition}:
\begin{equation}
\frac{1}{i\hbar}[\hat{A}, \hat{B}]\rightarrow\{A, B\}
\end{equation}
where $A, B$ are two functions of canonical variables corresponding to the quantum operators $\hat{A}, \hat{B}$ and $\{A, B\}$  denotes their usual Poisson Bracket. So, the classical counterpart of the relations (3) reads
 \begin{eqnarray}\label{E12}
\{ x_{i}, x_{j}\}= \alpha_{ij},\quad \{ x_{i}, p_{j}\}= \delta_{ij},
\quad \{ p_{i}, p_{j}\}= \beta_{ij}. \label{eq:poisson}
\end{eqnarray}
It is useful to note that $\hbar\alpha_{ij}$ must have dimension of  squared  length and in consequence $\alpha_{ij}$ has dimension of \emph{Time/Mass}.\\
In this work, we consider the case  $\beta_{ij}=0$,  then the NC classical mechanics described by  (5) reads:
 \begin{eqnarray}\label{E1}
\{ x_{i}, x_{j}\}= \alpha_{ij},\quad \{ x_{i}, p_{j}\}= \delta_{ij},
\quad \{ p_{i}, p_{j}\}= 0. \label{eq:poisson}
\end{eqnarray}
Now, by  the symplectic structure given in (6) and taking the Hamiltonian as
\begin{equation}
H=\frac{p_ip^i}{2m}+V(x_i),
\end{equation}
it is easily to show that the Hamilton's equations (1) become:
\begin{eqnarray}
&\dot x_{i}&=\frac{p_{i}}{m}+\alpha_{ij}\frac{\partial V}{\partial x^{j}}, \label{eq:,}\\
&\dot p_{i}&=-\frac{\partial V}{\partial x^{i}}.  \label{eq:h1}
\end{eqnarray}
In the next section, we  use the last equations to obtain the motion equations  for a test particle moving  in  central  potential $V=V(r)$.
\section{ The Equations of Motion}
In the  subsections below,  the equations of motion for the two following cases are presented:\\
1) A test charged particle moving in the NC space acted upon by a central force  generated by a static charged sphere.\\
2) A test charged particle moving in the usual space acted upon by  stationary electromagnetic field generated by a slowly rotating charged sphere.
\subsection{The Static Charged Sphere in the NC Space}
As discussed above, defining the NC classical mechanics equipped with symplectic structure (6), the motion equations,  for a test particle moving in a potential $V=V(x_i)$, are given by (8), (9). It is easy to show that   by eliminating the momentum variable from these equations, we get:
\begin{equation}\label{E11}
m\ddot x_{i}=-\frac{\partial V}{\partial x_{i}}+
m\alpha_{ij}\frac{\partial^{2 } V } { \partial x_{j}\partial x_{k}}
\dot x_{k }.
\end{equation}
Substituting the  Coulomb  potential $V(r)=\frac{qQ}{4\pi \varepsilon_0 r}$ into (\ref{E11}), gives
\begin{equation}\label{E2}
m\ddot x_{i}=\frac{x_{i}}{r}\frac{k }{r^{2}}+m\epsilon_{ijk}\dot x_{j} \Omega_{k}+m\epsilon_{ijk}x_{j}\dot
\Omega_{k},\label{eq:kepler}
\end{equation}
where  $k=\frac{qQ}{4\pi \varepsilon_0 }$ and
\begin{equation}\label{E22}
\Omega_{i}=\frac{-k}{r^{3}}\alpha_{i},
\end{equation}
has dimension of angular velocity.  As we will see, it  plays the role of rotation  and this role is even more highlighted, if we note that the second and third terms in the right--hand  side of (\ref{E2}) are similar to the \emph{virtual forces} which appear  in the  motion equations of a particle in the non--inertial (rotating) frames. But here, they actually are dynamical effects due to the noncommutativity. For example, the second term in the right--hand  side of (\ref{E2}) is just like the \emph{Coriolis force} appearing in the rotating system. This means that,  in the NC space, the test particle experiences a \emph{Quasi Coriolis force} in absence of  rotating frame. In other words, the noncommutativity  can produce dynamical effects  similar to  that which are produced by rotating frames, and this is due to the presence of quantity defined by (\ref{E22}) which can be write in the vectorial form as  $\vec{\Omega}=\frac{-k}{r^{3}}\vec{\alpha}$. Evidently, the (defined) angular velocity vector $\vec{\Omega}=(\Omega_{i})$  and  the (defined) NC vector $\vec{\alpha}=(\alpha_{i})$ are are parallel which induces  the physical rotation role for the NC vector.\\
In order to utilize  equations of motion (11), we set the noncommutativity to the $x-y$ plan, that is
\begin{center}$\vec{\alpha}=(\alpha_i)=(0, 0, \alpha)$,\end{center}
and hence the angular velocity  (12) becomes:
\begin{center}$\vec{\Omega}=(0, 0, \Omega= -k \alpha/r^3)$.\end{center}
By the above assumptions and writting  the motion equations  (\ref{E2}) in spherical coordinates $(r, \theta, \phi)$,  we get:
\begin{eqnarray}
m(\ddot r -r\dot \theta^{2}-r\dot \phi^{2}{\sin}^{2}\theta)=
-\frac{k}{r^2}+mr\Omega\dot \phi {\sin}^{2}\theta, \label{sphe1}\\
m \frac{d}{dt}(r^{2}\dot\theta)-mr^{2}\dot \phi^{2}{\sin}\theta{\cos}\theta=
-m r^{2}\Omega\dot \phi {\sin}\theta{\cos}\theta,\label{sphe2}\\
\frac{d}{dt}(mr^{2}\dot \phi {\sin}^{2}\theta)=mr{\sin}\theta
\frac{d}{dt}(r\Omega {\sin}\theta). \label{sphe3}
\end{eqnarray}
As usual, without losing  the generality, by taking the motion in the equatorial plane $\theta=\pi/2$, the  equations  (13)--(15) reduce to:
\begin{eqnarray}
&m(\ddot r -r\dot \phi^{2})=&\frac{k}{r^2}+mr\Omega\dot \phi=\frac{k}{r^2}-\frac{km\alpha}{r^2} \dot \phi, \\
&\frac{d}{dt}(mr^{2}\dot \phi )=&mr\frac{d}{dt}(r\Omega)=2mk\frac{\alpha}{r^2}\dot{r}, \label{eq:angular}
\end{eqnarray}
which by putting  $k=\frac{qQ}{4\pi \varepsilon_0 }$ and   simplification, read
\begin{eqnarray}
&\ddot r=& r\dot \phi^{2}- \frac{qQ \alpha}{4\pi \varepsilon_0 }  \frac{\dot \phi}{r^2}+  \frac{qQ}{4\pi \varepsilon_0 mr^2}   \label{eq:angular1},\\
&\ddot \phi=&-2\frac{\dot r}{r}\dot \phi+\frac{qQ \alpha}{2\pi \varepsilon_0 }  \frac{\dot{r}}{r^4}. \label{eq:angular}
\end{eqnarray}
The last two equations describe the motion of  test particle in an electrostatic field generated by a static charged  sphere  in the NC space.
\subsection{Spinning Charged Sphere in the Usual Space}
The classical electrodynamics Lagrangian (for a test particle of  charge $q$ and mass $m$), in terms of the scalar and vector potentials $(\Phi, \textbf{A})$, can be written as \cite{Jackson}:
\begin{equation}\label{D1}
 L= {\frac{1}{2}}mv^2 - q\Phi + q{\bf v} \cdot {\bf A}.
\end{equation}
Let the electromagnetic  potentials are due to an uniformly charged  rotating sphere with radius $R$ and charge
$Q$. The rotation  is specified by  constant angular velocity $\omega$ about the $z$-axis that
goes through the center of sphere. The scalar and vector potentials outside of  the sphere ($r>R$) are given by \cite{Jackson}:
\begin{equation}\label{D2}
\Phi=\frac{Q}{4\pi \varepsilon_0 r}, \hspace{1.5 cm} {\bf A}=\frac{Q\omega R^2}{5(4\pi\varepsilon_0)c^2}\frac{\sin\theta}{r^2}\hat{\phi},
\end{equation}
where the vector potential is in the magnetic dipole approximation form (that is $r\gg R$).
Substituting the potentials (21) into Lagrangian (\ref{D1}) gives:
\begin{equation}
L= {\frac{1}{2}}m(\dot r^2+r^2\dot \theta^2+r^2\dot\phi^2\sin^2\theta) - \frac{qQ}{4\pi\varepsilon_0 r} + \frac{qQ\omega R^2}{5(4\pi\varepsilon_0)c^2}\frac{{\sin}^{2}\theta}{r}\dot \phi,
\end{equation}
then, the Lagrange equations become:
\begin{eqnarray}
\ddot r =r\dot \theta^{2}+r\dot \phi^{2}{\sin}^{2}\theta
+\frac{qQ}{4\pi\varepsilon_0mr^2}- \frac{qQ\omega R^2}{5(4\pi\varepsilon_0)mc^2}\frac{{\sin}^{2}\theta}{r^2}\dot \phi, \label{sphe1}\\
\ddot \theta= -2\frac{\dot{r}}{r}\dot\theta+\dot \phi^{2}{\sin}\theta{\cos}\theta
+ \frac{qQ\omega R^2}{5(4\pi\varepsilon_0)mc^2} \frac{{\sin}2\theta}{r^3} \dot \phi,\label{sphe2}\\
\ddot \phi=-2\frac{\dot{r}}{r}\dot\phi-2\dot \theta \dot \phi\cot\theta-\frac{2qQ\omega R^2}{5(4\pi\varepsilon_0)mc^2}  \frac{\cot\theta}{r^3}\dot \theta+\frac{qQ\omega R^2}{5(4\pi\varepsilon_0)mc^2} \frac{\dot r}{r^4}, \label{sphe3}
\end{eqnarray}
which for ($\theta=\pi/2$) are reduced to
\begin{eqnarray}
\ddot r =r\dot \phi^{2}-\frac{qQ\omega R^2}{5(4\pi\varepsilon_0)mc^2}\frac{\dot \phi}{r^2}
+\frac{qQ}{4\pi\varepsilon_0 m r^2}\label{B}, \\
\ddot \phi=-2\frac{\dot{r}}{r}\dot\phi  +\frac{qQ\omega R^2}{5(4\pi\varepsilon_0)mc^2} \frac{\dot r}{r^4}\label{A}.
\end{eqnarray}
The last two equations describe  the motion of test particle in an  electromagnetic  field generated by a rotating charged sphere (as a magnetic dipole) in the usual space. In the next section,  the two sets  equations given in (18, 19)   and (26, 27) are employed to obtain an equivalence relation between the noncommutativity and the intrinsic magnetic  field (magnetic dipole effect).
\section{Equivalence Between Noncommutativity and Intrinsic Magnetic Field}
\subsection{Equivalence Relation}
In order to show equivalency between effects of  noncommutativity   and the intrinsic magnetic field, we should compare the two sets of motion equations, that is (18, 19)   and (26, 27). The similarity between the  two sets equations makes the comparison   easy. But before the comparison, let us remember that the  set (26, 27)  are subject  to the dipole approximation condition $r\gg R$. This condition also must be met for the set (18, 19)  and therefore it is legitimate to ignore the terms of the $1/r^4$  from the equations  (19) and (27). By neglecting the  fourth order term  and comparing the second terms in equations  (18), (26), we obtain:
\begin{equation}\label{Equivalence1}
k\alpha\equiv\frac{qQ\omega R^2}{5(4\pi\varepsilon_0)mc^2},
\end{equation}
which by substituting $k=qQ/4\pi\varepsilon_0$ becomes
\begin{equation}\label{Equivalence}
\alpha\equiv \frac{R^2}{5mc^2}\omega \propto \omega.
\end{equation}
We call the relation (29), the equivalence relation which can also be written in the vectorial  form
\begin{equation}
\vec{\alpha}\equiv \frac{R^2}{5mc^2}\vec{\omega}\propto \vec{\omega}.
\end{equation}
In connection with the equivalency relation (29), the following notes may be useful:\\
1- The result ($\alpha \propto \omega$) indicates that the NC effect ($\alpha$ considered as representative parameter of noncommutativity) can play the role of  rotation ($\omega$ considered as representative parameter of rotation). Recall that $\alpha$ is a geometrostatic property of the NC space and $\omega$ is a kinematical property  in the usual space. So, the geometrostatic property of the NC space manifests itself as a  physical (kinematical) effect.\\
2- A corollary which draws  attention is  ($\alpha \propto 1/m$), the NC parameter is also proportion to the  inversed mass. This means that the noncommutativity may be responsible for  physical effects caused by the particle mass. It is necessary to note that,  the mass of test particles in the usual and NC spaces are taken to be equal, however  the  mass and the other quantities  which appear in the right hand side of equation (29) are  of the usual space.\\
3- $\omega$ and $m$ are quantities with  a purely physical character  whereas $\alpha$ has   a geometric identity. Therefor, in this respect, relation (29) can be important because it demonstrates the interplay  between geometry and physics. In other words, one might argue,  the source of the physical quantities (effects) may be attributed to geometry or vise versa.\\
4- The relation (28) is not an equation to determine the constituent quantities, say $\alpha$, as it may appear in relation (29).
Because, as mentioned, the left hand side of this equation comes from the NC space and the right hand comes from the usual space. Indeed, when the relation (28) is established,  the equations of motion in both spaces (NC and usual) are the same and equation (28) is to compare the values of the related quantities in these two spaces, specially  $\alpha$ and $\omega$. Another simple explanation  can be stated as follows:
if the equivalence relation  (28) is hold, then a NC space can be constructed  with the NC parameter given by relation (29), as result the motion equations of  test  particle  in  both  spaces (NC and usual) become the same.\\
5- In obtaining the  motion equations (18), (19) and consequently  the equivalence relation (28), the radius of charged (static) sphere (in the NC space)  does not play a role. Hence, it can be, for example,  a point charge or any other static spherical symmetric distribution of charges.\\
6-  It is true that the relation (29) was obtained based on dipole approximation $r\gg R$ (ignoring terms of order $r^{-4}$ in equations $(\ref{eq:angular})$ and $(\ref{A})$), however without such an approximation, we can also get the same result. To illustrate this, we  note that, the constant coefficients of the second terms, in equations (\ref{eq:angular1}) and (\ref{eq:angular}), differ by a factor of two and since the values of NC parameter $\alpha$ are (typically) very  small\rlap,\footnote{For example, it has been shown that the NC parameter $\alpha$ is order of $10^{-62} m^2$\cite{Mirza} or $10^{-58} m^2$\cite{Romero}.} hence a factor of two in the NC parameter can not make a considerable impact on the main and final results. In other words, the constant coefficients of the second terms in equations (18) and (19) can be approximately equal just like the constant coefficients of the second terms in equations (26) and (27).
\begin{figure}
\centering
  \includegraphics[width=5.5cm]{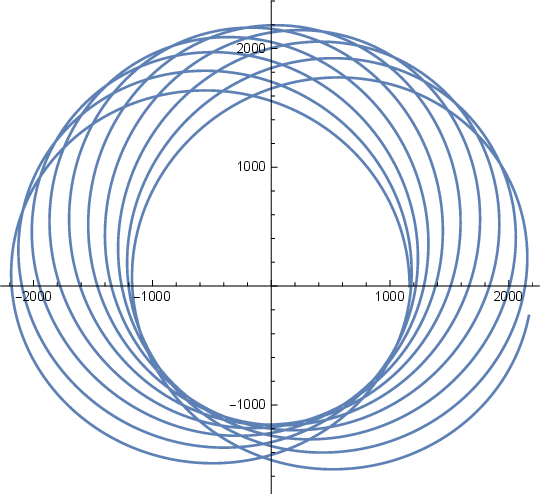}\hspace{1cm}
  \includegraphics[width=5.5cm]{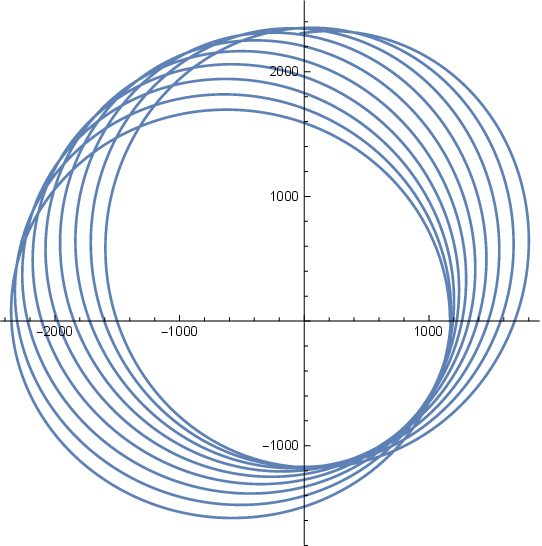}
  \caption{The trajectories of a charged particle in the equatorial plane  around charged sphere  in the noncommutative space (left panel) and usual space (right panel). The electric charges of the particle and sphere are of  opposite sign (attractive case).}
\end{figure}
\begin{figure}
\centering
  \includegraphics[width=5.5cm]{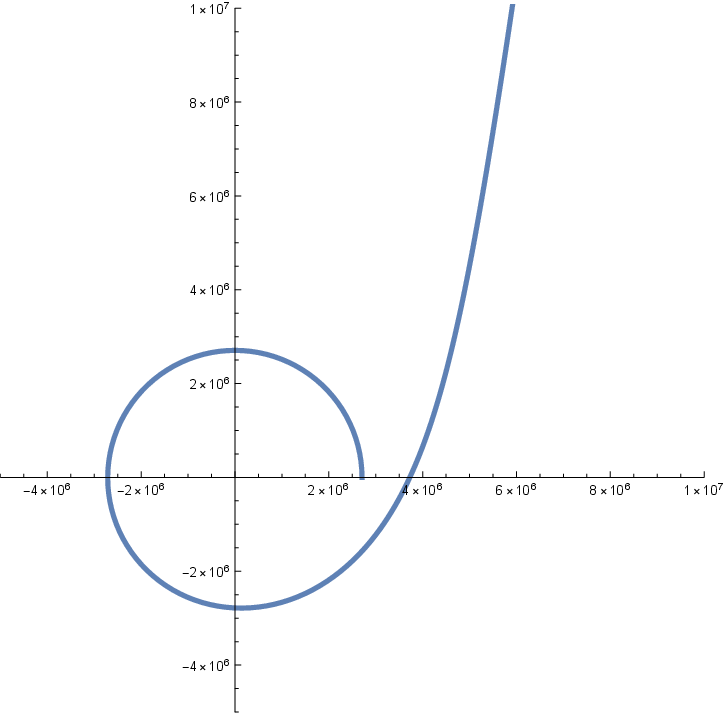}\hspace{1cm}
  \includegraphics[width=5.5cm]{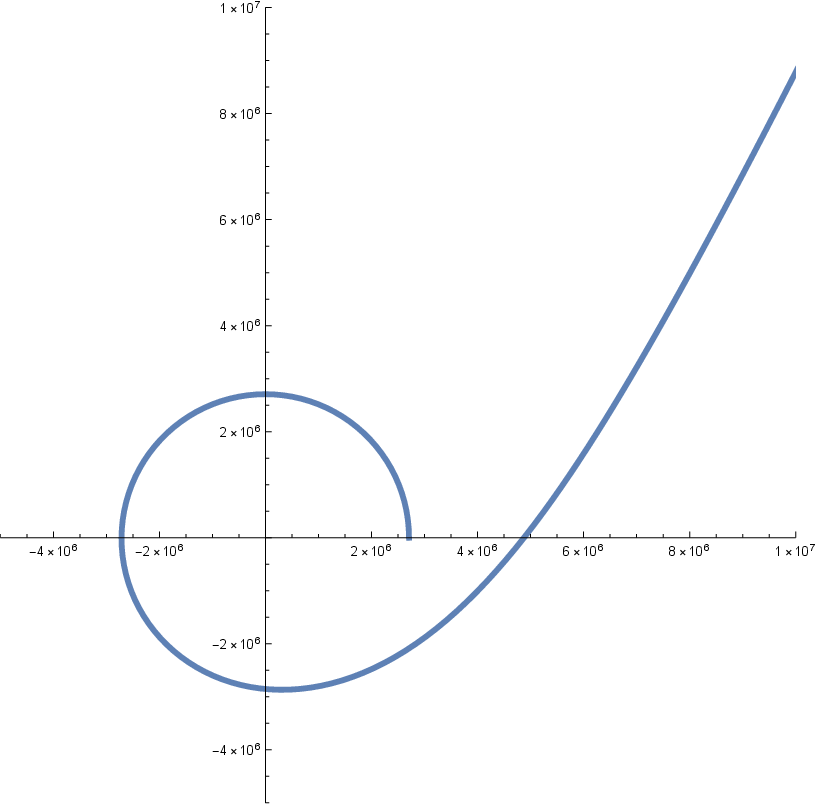}
  \caption{The trajectories of a charged particle in the equatorial plane around charged sphere  in the noncommutative space (left panel) and usual space (right panel). The electric charges of the particle and sphere have the same sign (repulsive case).}
\end{figure}
\subsection{Graphical Support}
So far, we have obtained the equations of motion of a charged particle in two usual  and NC spaces, and by comparing the equations of motion, we have reached the equivalence relation (29), but we have not talked about the type of particle motion, means that in electromagnetism, the  interaction  between  charged particles includes two types of attraction and repulsion.   On the other hand, the equivalence relation  requires that the motion of the particle in two spaces is qualitatively similar\rlap.\footnote{Because it is obtained from  similarity of the equations of motion.} In this subsection, we want to plot the   equatorial paths for a test charged particle whose motion equation are given by (18),(19) (NC space) and  (26),(27) (usual space). It can be observed that ploting  the paths indictes a goodness support for the equivalence relation. The paths are plotted for the following two cases:\\
1) The charge of the teast particle $q$ and charge of the sphere $Q$ have  the opposite sign (attractive case).\\
2) The charge of the teast particle $q$ and charge of the sphere $Q$ have  the same sign (repulsive case).\\
To plot the paths, first note that the two sets of motion equations ((18),(19))  and  ((26),(27)) are in   polar coordinates $(r, \phi)$ which are related to Cartesian  coordinates by ($x(t)=r(t) \cos \phi(t), y=r(t)\sin \phi(t)$).  As result, by removing the time between the Cartesian coordinates, the path of the particle in the Cartesian (equatorial) plane can be depicted. For the two sets of motion equations ((18), (19))  and  ((26) and (27)), it can be done   by the appropriate  numerical computational codes. The graphs shown in figures 1, 2 are the outpus of implementing the corresponding codes for the two set  equations.\\
Let's remember that equations ((18), (19)) describe the motion of a charged particle under Coulomb's potential (due to a static charged sphere) in the NC space and equations ((26), (27)) describe the motion of a charged particle under electromagnetic field (due to a charged rotating sphere) in the usual space.\\
The graph in  Fig. 1-left (right) panel  shows the path of the particle for the attractive case in the NC (usual) space and the graphs in  Fig. 2- left (right) panel  show the path of the particle for repulsive case in the NC (usual) space.\\
In all the graphs of  figures 1, 2, the numerical values for the Coulomb's contant and speed of light are  $1/4\pi \varepsilon_0=9 \times 10^9$ and  $c=1$.\\
For  ease of availability, the names and values of the parameters in two sets of motion  equations (18, 19)  and (26, 27) are given below.\\
\textbf{Names}:\\
$Q=$ Charge of the central sphere (static or  rotating).\\
$q, m=$ Charge  and mass of the test particle.\\
$\alpha=$ NC parameter.\\
$\omega=$ Angular velocity of the rotating sphere.\\
$R=$ Radius of the rotating sphere.\vspace{1mm}\\
\textbf{Numerical values}:\\
\begin{center} \textbf{Fig. 1 (attractive case)}:\end{center}
Left panel (NC space): The numerical values for the parameters in the motion equations (18), (19) are:
\begin{center} $Q=\frac{10^{14}}{9},  m=4.5 \times 10^{25},  q=-2.1\times 10^{13}, \alpha=10^{-28}.$\end{center}
Right panel (usual space): The numerical values for the parameters  in the motion equations (26), (27) are:
\begin{center}  $Q=\frac{10^{15}}{9}, m=4.5 \times 10^{25} ,  q=-2.1\times 10^{12}, \omega=10^{-5}, R=47.43.$\end{center}
\begin{center} \textbf{Fig. 2 (repulsiveive case)}:\end{center}
Left panel (NC space): The numerical values for the parameters  in the motion equations (18), (19) are:
\begin{center}  $Q=10^{21},  m=4.5 \times 10^{25}, q=2.1\times 10^{15}, \alpha=10^{-26}.$\end{center}
Right panel (usual space): The numerical values for the parameters  in the motion equations (26), (27) are:
\begin{center}  $Q=10^{21},  m=4 \times 10^{20}, q=1.9 \times 10^{10}, \omega=10^{-4}, R=150.$\end{center}
It should be noted that the  large values for electric charges is due to the existence of such charges in the physical world. In article \cite{Neslusan}, it is shown that   the electric charge inside a stellar sphere  is linearly proportional to mass  inside the sphere. For exapmle, for a typical star as sun, $Q$ turns out to be $1.5\times 10^{28}$ (Gaussian unite).  In addition, the large values of the electric charges ensure the negligibility of the gravitational interaction against the electric interaction.
\subsection{Revisited Equivalence Relation}
In electrostatics, the simplest field structure is the electric monopole (point electric charge), while in
magnetostatics, the elementary field configuration found in the nature is the magnetic dipole (the most basic type of magnetic multipoles). Indeed, we know that,  the magnetic field produced by a circuit\footnote{The  most fundamental object that posses the magnetic moment  is a current loop as simplest circuit.}  far away from its location does not depend on  specific geometrical shape of the circuit but only of magnetic moment. This is the dipole approximation that makes all circuits with the same magnetic moment  produce the same  magnetic field.

As we see,  the  equivalence relation (28)  includes the angular velocity $\omega$  and radius of the rotating  sphere $R$,  however,  rotating charged sphere  is only one of the types of current distributions for which the dipole approximation is applicable. On the other hand, one of  the main purpose of this work is to show equivalence relation between noncommutativity and magnetic dipole effects, so,  it is appropriate to describe a relationship between noncommutativity  and the magnetic moment of the circut. In other words, the desired  (equivalence) relation should be independent from the shape  of the magnetic dipole, for example, the raduis of the rotating sphere $R$. Therefore, since all circuits with the same magnetic moment  produce
the same  magnetic field, we can  obtain a general relation for an arbitrary magnetic dipole. To do this, first notice  that the magnetic  moment of the rotating charged sphere is:
\begin{center}$\vec{\mu}= \frac{Q R^2}{5}\omega \hat{z}=\frac{Q R^2}{5}\vec{\omega},$\end{center}
which by making use of   relation  (30), it is easy to obtain:
\begin{equation}
\vec{\alpha}\equiv \frac{\vec{\mu}}{Qmc^2}\propto \vec{\mu} \hspace{1cm} Revisited \hspace{2mm} Equivalence\hspace{2mm}Relation,
\end{equation}
where now the magnetic moment $\vec{\mu}$ can have any circute current  source. In fact,  relation (31) is a more fundamental expression of the equivalency concept. It directly  demonstrates the relation between the fundamental source of the magnetic field and the noncommutativity effect. But, one should be a little careful with this relation, since, it is not a relation to determine the magnetic moment from the NC parameter or vise versa. It is a relation to express the equivalency between geometrostatic effects of the (NC) space and a physical effects (magnetic dipole) in the usual space. In a clearer expression, we can say that, in the NC space, despite the absence of a magnetic dipole, there is a similar effect caused by noncommutativity whose equivalent magnetic  moment (by relation (31)) can be written as
\begin{equation}
\vec{\mu}_{nc}\equiv Qmc^{2}\vec{\alpha},
\end{equation}
where  $\vec{\mu}_{nc}$ stands for the equivalent magnetic moment or NC companion magnetic moment (defined in the NC space). The last equation allows us to go further and define the equivalent magnetic field  or NC companion magnetic field as follows:\\
In the usual space,  the magnetic vector potential $\vec{A}$  and
the magnetic field $\vec{B}$ created by the magnetic dipole $\vec{\mu}$ are given by:
\begin{center}$\vec{A}=\frac{1}{c}\left(\frac{\vec{\mu}\times\vec{r}}{r^3}\right),  \hspace{1 cm}\textrm{(Gaussian units)}$\end{center}
and
\begin{equation}
\vec{B}=\vec{B}(\vec{\mu}, \vec{r})=\bigtriangledown \times \vec{A} =\frac{1}{c}\left(\frac{3(\vec{\mu}\cdot \vec{r})\vec{r}}{r^5}-\frac{\vec{\mu}}{r^3}\right).
\end{equation}
Now, by inserting the NC magnetic moment (32) into equation (33), the NC companion magnetic field $\vec{B}_{nc}$ can be defined as:
\begin{equation}
\vec{B}_{nc}=\vec{B}(\vec{\mu}_{nc}, \vec{r})=Qmc\left(\frac{3(\vec{\alpha}\cdot \vec{r})\vec{r}}{r^5}-\frac{\vec{\alpha}}{r^3}\right).
\end{equation}
From the physical point of view, the last formula can  be interpreted as follows:
In presence of a  static sphere of electric charge $Q$ in the NC space, a moving test particle of mass $m$, charge  $q$ and velocity  $\vec{v}$ experiences a quasi magnetic force as
\begin{equation}
\vec{F_{nc}}=(q/c) \vec{v}\times \vec{B}_{nc}=m(qQ) \vec{v}\times     \left(\frac{3(\vec{\alpha}\cdot \vec{r})\vec{r}}{r^5}-\frac{\vec{\alpha}}{r^3}\right).
\end{equation}
Note that the NC magnetic field  (34) depends on the mass of the test particle (contrary to the magnetic field in usual space), means that, the NC magnetic dipole has also interact with the physical  mass. Also, the quasi magnetic  force (35) is  proportional to the mass (again contrary to the usual space), so the NC acceleration\footnote{The acceleration of the test particle.}    can be obtained as:
\begin{equation}
 \vec{a_{nc}}=\vec{F_{nc}}/m=(qQ) \vec{v}\times \left(\frac{3(\vec{\alpha}\cdot \vec{r})\vec{r}}{r^5}-\frac{\vec{\alpha}}{r^3}\right).
\end{equation}
In the end of this section, it would be best to make a simple example to exploit the above discussion.\\
For circular motion in the $x-y$ plan, the formulas (34) and (36) reduce to:\footnote{In this motion, the NC vector $\vec{\alpha}$ and velocity $\vec{v}$ are perpendicular to position vector $\vec{r}$.}
\begin{center} $\vec{B}_{nc}=-Qmc\left(\frac{\vec{\alpha}}{r^3}\right)$\hspace{3mm}  and \hspace{3mm} $\vec{a_{nc}}=-  \alpha \frac{q Q v}{r^4} \vec{r},$ \end{center}
which by equating the last term with the centrifugal acceleration ($-r \omega^2$),  the NC angular velocity can be obtained  as $\omega^2=\omega^2_{nc}=  \alpha \frac{qQ v}{r^4}$. On the other hand, the cyclotron frequency of this circular motion is $\omega_{c}=  \frac{q B_{nc}}{m}=\alpha\frac{Qqc}{r^3}$ whose square must be equal to  $\omega^2_{nc}$, therefore, it can be easily resulted that \begin{center} $Qqc^2 \alpha=v r^2$.\end{center}  Note  that all the quantities in the last equation are in the NC space, contrary to equivalency relation (28) whose left and right  hand sides  are  for NC space and  usual space, respectively. Therefore it can be consider as relation between the radius and velocity of the circular motion with the values of NC parameter. In simpler terms, for example, if $q=Q$ and for a given   velocity $v=c/n$ which $n$ can be a positive integer, the  radius of circular motion (in natural unites $c=1$) can be obtained as $r=q\sqrt{n \alpha}$. This formula can be also used in reverse to measure the NC parameter in an empirical experiment.
\section{General Discussion}
The presence of  the  magnetic field in a region of space depends on the movement of electric charge relative to the observer. In other words, two necessary components  for the emergence of the magnetic field, are the entity called electric charge and  the motion of charge (electric current). Put in a short sentence: without a moving charge, a magnetic field can never exists. There are two known types of electric current, the translation and rotational currents. The magnetic field generated by the first type can be eliminated in the rest frame of the moving charge, and that is why it is sometimes called extrinsic magnetic field. The magnetic field generated by the second type comes directly from intrinsic (spin) motion of the charge distribution\rlap,\footnote{The charge distribution has a proper angular momentum.} and most likely the reason why it is called  intrinsic magnetic field. Contrary to the extrinsic  field, the intrinsic type can not be eliminated by the coordinates transformation and  this is the main and  substantive distinction between them. The  magnetic dipole field  is a familiar example of the second type. The magnetic dipole is one of the most important concepts in magnetostatics and  is a fundamental object when the interactions between matter and magnetic fields are considered.

One of the main objectives of this article is to show  that  the effects of  a magnetic dipole (as an intrinsic type) can be attributed to geometro--static properties of the space--time. This objective can be put into the following statement:  the physical effects  similar to that of generated by  a magnetic dipole can be generated  when the main component to emerge of the magnetic field (moving electric charge) is absent. This is what we call equivalency between NC effects and intrinsic magnetic field.

To demonstrate equivalency,  we first write down the equations of motion for a test particle in the both NC and usual spaces. In the NC space a spherically static distribution of charge  generates an electrostatic field. In the usual space, the electromagnetic field is generated by an uniformly rotating charged sphere subjected to the dipole approximation. By  comparing the equations of motion in the both spaces, it turns out that the role of the noncommutativity (in the NC space) is similar to the rotational effect of the charged sphere (in the usual space). In the other words,  in the NC space  containing  a static charged sphere, there is an magnetic field  similar  to a magnetic  field caused by a dipole. It should be recall that our approach in achieving the  result is based on comparing the equations of motion. In the usual space, the motion equations are obtained from the Lagrangian formalism and in the NC space, they are obtained from the Hamiltonian one. The latter formalism in turn is strongly dependent on the  symplectic structure defined on the phase space manifold. In geometric language, the results is directly dependent on the particular form chosen for the geometrical structure of the manifold.\\
So far we have tried to attribute magnetic (dipole) field to  geometric property (noncommutativity) of the space--time. If we succeed in doing so, it is not wrong to gather that there is an interplay between geometry and physics or dynamics. Indeed, there is an another straightforward approach \cite{Isidro} which shows that the phase space of the electromagnetic field has NC structure. Furthermore, it is shown that, for a
classical charged particle in an electromagnetic field,  under the special condition and by a new definition of the (generalized) Poisson bracket,  the phase space variables no longer commute in the sense of the new bracket. The new bracket describes the motion equations with respect to the new Hamiltonian.
This approach is based on the Dirac theory \cite{Dirac} which is the extension of the Hamiltonian formalism for dynamical system including constraints. Also, it has been indicated that, in the presence of an external background magnetic field, the NC coordinates can naturally be introduced \cite{Seiberg}, that is, the (quantum) commutator of the phase space coordinates are no longer the Poisson commutator.
\subsection{A short scape into  Geometrodynamics}
At this point, it will not be irrelevant to give a brief talk about the \emph{Geometrodynamics}, because, the above discussions and results resemble apparently the situation similar to what encounters in geometrodynamics.\\
As we know, in the usual relativity, the matter (particles and fields) distributions are considered as  physical entities immersed in geometry and are responsible for the  space--time curvature (entities independently of geometry, somethings external with respect to the space). In geometrodynamics, on the contrary, the curved space--time is assumed to be free from the matter distributions (what is sometime called empty space). That is the curvature is not due to  masses and fields, but they are   considered as the manifestations of geometry. The most familiar example is  the gravity which is caused by the geometric property of space--time, that is curvature. The another example can also be familiar from Quantum Geometrodynamics in which a quantum potential (Bohmian one) may be considered as geometrodynamic entity \cite{Licata}.\\
Therefor, in a general respective, the geometrization of  physical phenomena can be regarded as the link point between the present work and geometrodynamics. But, there is a small difference: In geometrodynamics, all physical processes  are viewed as manifestations of  the (curvature) geometry, but, in this work, a  geometric property (noncommutativity)  manifests itself as a  physical effect (magnetic dipole).
\subsection{Intrinsic and Extrinsic Equivalencies}
In this subsection, we want to draw the attention of the reader to the fact that the equivalence relation (28) (which we call it intrinsic equivalency) is obtained based  on the commutation relations (5) (subjected to the condition $\theta_{ij}\neq 0, \beta_{ij}= 0$).
As mentioned above, the  equivalency between NC effects and extrinsic magnetic field (which we call it Extrinsic Equivalency) has been shown\cite{Djamai}. This equivalency has been also obtained based on commutation relations (2), but  subjected to the conditions  $\theta_{ij}=0, \beta_{ij}\neq 0$.

Note that in the intrinsic (extrinsic) equivalency, the NC parameter corresponding to the coordinate (momentum) sector of the phase space is present and the other is absent, but both have one thing in common, which is   they  play the role of movement of the charge distributions. In the other words,  the NC parameter corresponding to the coordinate (momentum) sector play the role of the rotational (translational) motion of the (static) charge  distributions  in the NC space. In the other words, we deal with a physical effect (magnetic field) whose existence depends on the motion (of  the electric charges), but here, it emerges as a manifestation a \emph{geometrostatic} property of the space--time, that is noncommutativity.\\
Since the physical and geometrical phenomena of our model are independent of time, it may be useful to express the results from this point of view as follows: \begin{center} A \emph{gemetrostatic} property of the space can produce a \emph{magnetostatic}  effect.\end{center}
From another perspective, we note that the noncommutatvity is a micro--scale geometric property and the magnetic field is a macro--physical effect depend on  motion, then one can say: \begin{center}A  \emph{micro--static} geometric property of the space   can beget a \emph{macro--motional} physical effect.\end{center}
\section{Conclusions}
Perhaps it could be said that the science of magnetism is part of the most fascinated sciences and its history go back to the dawn of the human race.
The magnetism has preoccupied the minds of many researchers for many decades to obtain the  research accomplishments of the different magnetic phenomena in theoretical and experimental  scopes.

On the other hand, all physical processes (including magnetic phenomena) occur in space--time which in classical picture has a continuous structure, however, quantum picture  do not allow space--time divisions below a certain length (scale). Therefore, the identification  of the geometric structure
of space--time and clarification of  its properties (in micro and macro scales) can   play a most important role in analyze of physical processes.
The NC geometry describes the space--time structure at micro scales and in fact it arises from the intrinsic properties of space--time  which  is a framework for the physical events. From this point of view, the effects of accompanying noncommutativity may have a relevance. As a result, it could be interesting to study within this area  and finding answer to question like:\\
Can physical quantities (as   diploe magnetic field)  be attributed to the geometrical properties of space--time?, or  in reverse:\\
Can geometrical properties produce physical effects as magnetic dipole?\\
The above questions are posed in the light of the equivalence relation (28) and from this  perspective, which is intended for the present work, they can also be asked as follows:\\
Is it possible to exist or manifest a physical phenomenon in the space--time when   its physical source is absent?

At the point of reasoning and methods of the present work, we can answer the above questions in the affirmative. This means that, affectations corresponding to a physical phenomenon can be emerged when its physical sources   are not present in the surrounding space--time. Instead of the physical  sources,  the micro scale properties\footnote{By micro scale properties, we mean the noncommutativity.} of the space--time could be (an alternative)  responsible for the affectations. Accordingly, one can say a \emph{geometrostatic} property (as noncommutativity)  can play the role of the source for a magnetic dipole when  its main component (electric current)  is absent.\\
As a last word, we remember that the results obtained in this study are fully dependent on the choice of  particular NC  geometry of  the phase space which manifests itself in peculiar form of the symplectic structure. Obviously, the other choices of symplectic structure may lead to  different results, that is the noncommutativity  might be able to play the role of other physical or dynamical effects. This can be  motivation and idea for future researches in the  area.

%
\end{document}